\documentclass[12pt]{article}
\usepackage{ametsoc}
\usepackage{natbib}
\usepackage[utf8]{inputenc}
\usepackage{pdflscape}
\usepackage{rotating}
\usepackage{soul}
\newcommand{\citepossess}[1]{\citeauthor{#1}'s \citeyearpar{#1}}

\newcommand{\lna}[1]{{\color{green}#1}}

\newcommand{\myabstract}{The possibility of constructing Lorenz's concept of available potential energy (APE) from a local principle has been known for some time, but has received very little attention so far. Yet, the local APE framework offers the advantage of providing a positive definite local form of potential energy, which like kinetic energy can be transported, converted, and created/dissipated locally. In contrast to Lorenz’s definition, which relies on the exact from of potential energy, the local APE theory uses the particular form of potential energy appropriate to the approximations considered. In this paper, this idea is illustrated for the dry hydrostatic primitive equations, whose relevant form of potential energy is the specific enthalpy. The local APE density is non-quadratic in general, but can nevertheless be partitioned exactly into mean and eddy components regardless of the Reynolds averaging operator used.
 
This paper introduces a new form of the local APE that is easily computable from atmospheric datasets. The advantages of using the local APE over the classical Lorenz APE are highlighted. The paper also presents the first calculation of the three-dimensional local APE in observation-based atmospheric data.  Finally, it illustrates how the eddy and mean components of the local APE can be used to study regional and temporal variability in the large-scale circulation. It is revealed that advection from high latitudes is necessary to supply APE into the storm track regions, and that Greenland and Ross Sea, which have suffered from rapid land ice and sea ice loss in recent decades, are particularly susceptible to APE variability.}

\begin{document}

%
%
\title{\textbf{\large{On the Local View of Atmospheric Available Potential Energy}}}
%
%
\author{\textsc{Lenka Novak,}
				\thanks{\textit{Corresponding author address:} 
				Lenka Novak, Department of Meteorology, 
				University of Reading, P.O. Box 243, Reading, RG6 6BB, United Kingdom. 
				\newline{E-mail: l.novakova@reading.ac.uk}}\quad\textsc{R\'emi Tailleux}\\
\textit{\footnotesize{Department of Meteorology, University of Reading, Reading, United Kingdom}}
\and 
}
%
\ifthenelse{\boolean{dc}}
{
\twocolumn[
\begin{@twocolumnfalse}
\amstitle

\begin{center}
\begin{minipage}{13.0cm}
\begin{abstract}
	\myabstract
	\newline
	\begin{center}
		\rule{38mm}{0.2mm}
	\end{center}
\end{abstract}
\end{minipage}
\end{center}
\end{@twocolumnfalse}
]
}
{
\amstitle
\begin{abstract}
\myabstract
\end{abstract}
\newpage
}

\section{Introduction}

The stored potential energy that is available to fuel global circulation and the kinetic energy that quantifies that circulation are two key diagnostics that summarize the global state of the dynamical and thermodynamic properties of the atmosphere and oceans. As a result these energies and the conversions between them are commonly diagnosed in global climate change and model verification studies (e.g. \citealp{ogormanschneider08,mbengueschneider17}).

It has long been recognized that energy budgets are only useful if the potential energy (PE) is partitioned into its available (APE) and background (PE$_r$) components, following \citepossess{lorenz55} pioneering work. Indeed, this is because there is often no direct correspondence between variations of potential energy and variations of kinetic energy, as in the case of the 'cooling paradox', whereby cooling results in the creation of kinetic energy despite being a net sink of potential energy. In contrast, variations in APE are a much better predictor of variations in kinetic energy. However, a major difficulty with Lorenz's APE is that it is only defined in a global and volume-integrated sense. With an increasing emphasis of climate change research on regional variability in high resolution climate models, there is an increasing need for locally definable diagnostics that can summarize large amounts of data. 

While the local character of kinetic energy is already well established and widely used, the possibility to define APE from a local principle remains poorly known, despite it being proved over 30 years ago in two seminal papers by \cite{andrews81} and \cite{hollidaymcintyre81} for a compressible non-hydrostatic and an incompressible fluid respectively. This paper advocates the use of the local APE framework and demonstrates its applicability to the discussion of various aspects of atmospheric energetics in the context of the hydrostatic primitive equations for a dry atmosphere.

The available potential energy was first defined formally by \citet{lorenz55} as the difference between the total global potential energy of the actual state of the atmosphere and its adiabatically rearranged reference state. For a dry hydrostatic atmosphere viewed as a perfect gas and in absence of orography, one possible exact expression for Lorenz APE is as follows:  
\begin{equation}
\begin{split}
\overline{APE_{\text{Lor}}}
=\frac{c_p}{g p_0^{\kappa}}\frac{1}{(1+\kappa)} \int_0^{\infty} \overline{p^{\kappa+1}}-\overline{p}^{\kappa+1} d\theta , \label{eq1}
\end{split}
\end{equation}
where the bar denotes averaging over isentropic surfaces, $c_ p$ is the specific heat capacity at constant pressure, $g$ the gravitational acceleration, $p$ the pressure with $p_0$ being its mean surface value, $\kappa=R/c_p$ where $R$ is the gas constant for dry air, and $\theta$ is the potential temperature. Although Eq. \ref{eq1} is exact, it is generally regarded as computationally impractical so that, in practice, a majority of APE studies have resorted to using the so-called quasi-geostrophic (QG) approximation, which depends on the temperature variance on isobaric surfaces divided by static stability:
\begin{equation}
\begin{split}
\overline{APE_{\text{Lor}}}
\approx \frac{1}{2} \frac{\kappa c_p}{g p_0^{\kappa}} \int^{p_0}_{0} p^{-(1-\kappa)}\left(- \frac{\partial\overline{\theta}}{\partial p}\right)^{-1}\overline{\theta^{\prime 2}} dp, \label{Eq2}
\end{split}
\end{equation}
with the bar here denoting an average over isobaric surfaces (more detail on this derivation can be found, for example, in \citealp{grotjahn93}). 

This definition is a common diagnostic for global characteristics of APE in climate models and observation-based data \citep[e.g.,][]{huetal04,schneiderwalker08,ogormanschneider08,hernadezdeckersvonstorch10,veigaetal13}, as well as for studying the evolution of individual eddies in idealized lifecycle experiments \citep{simmonshoskins78}.
However, a limitation of the quadratic approximation \ref{Eq2} is that it assumes a small departure from the reference state, which may become potentially very inaccurate in areas of substantial mixing and rapidly varying static stability, such as the mid-latitude storm tracks \citep[e.g.,][]{hollidaymcintyre81}. Furthermore, Lorenz's definition is a global one and hence obscures regional variability and can lead to misleading results. For example, \citet{novaketalIR} showed that meridionally confined storm tracks exhibit a spatially complex thermal equilibration, which can be translated to a local APE decrease but a global increase globally as a response to polar cooling.  Such spatially complex responses cannot be captured by Lorenz's global APE definition. 

So far, most attempts at seeking a local view of energetics have relied on 'localizing' Lorenz APE by assuming that it is physically meaningful to study the spatial distribution of the integrand of Eq. \ref{Eq2} \citep[e.g.,][]{AAlietal07} or Eq. \ref{eq1} \citep[e.g.,][]{ahbecaldaira17}. Although these approaches appear to yield plausible results, it goes without saying that it would be far more satisfactory to base such analyses directly from a truly local definition of APE. Other attempts of using the Lorenz energetics locally include spatial integrations over a local domain of an open system which is embedded within a closed global system (\citealp{johnson70}). Though an exact framework, the precise spatial distribution of the various energy, conversion and transport terms is still obscure and the need for a different formulation of a local definition is apparent.

One important concept introduced as an attempt to resolve the difficulties associated with the global character of Lorenz APE is that of ``exergy''. In the context of atmospheric and oceanic sciences, exergy can be viewed as essentially measuring the departure of a system from its thermodynamic and mechanical equilibria. Such equilibria can be identified by defining an isothermal reference state, which was advocated by many (e.g., \citealp{dutton73}; \citealp{pearce78}; \citealp{blackburn83}; \citealp{karlsson90}). Although exergy is appealing due to its simplicity and local character, it is nevertheless fundamentally different and in general excessively larger than Lorenz APE (as stressed by \citealp{tailleux13b}). This is a result of the exergy depending on the system being brought to a maximum state of entropy when computing the reference state, whereas Lorenz's APE depends on the system being adiabatically rearranged whilst conserving entropy. It means that in contrast to APE, the total exergy of a system includes a large chunk of the background potential energy $PE_r$, which is a 'heat-like' form of potential energy and hence strongly constrained by the second law of thermodynamics.

So far, the only satisfactory approaches to construct Lorenz APE from a local principle appears to those stemming from the two studies by \citet{hollidaymcintyre81} for an incompressible fluid, and \citet{andrews81} for a fully compressible stratified one-component fluid. For both kinds of fluid, the authors were able to construct a locally defined positive form of potential energy density that can be interpreted as the work necessary to bring a parcel from its reference position $Z_r$ to its actual position $Z$. Thus in the case of an incompressible fluid, the APE density (in units J kg$^{-1}$) takes the following simple form:

\begin{equation}
E_a=\frac{g}{\rho_0}\int^Z_{Z_r} (\rho-\rho_r(Z',t)) dZ^{\prime}, \label{EaEq}
\end{equation}
where $\rho$ is the density and the subscript $r$ indicates the reference variables. $E_a$ is positive definite and by assuming a small departure from the reference state, it can be reduced to the APE of small amplitude internal waves (\citealp{hollidaymcintyre81}): $1/2 N^2 \zeta^2$, where $N$ is the buoyancy frequency and $\zeta = Z-Z_r$ is the vertical displacement from the reference position. 

Aside from its local nature, other advantages of this formulation over Lorenz APE are that it is exact, valid for finite amplitude departures from the reference position, \st{and} computationally easy to implement, and definable for a wider classes of reference states (such as horizontally- or isobarically-averaged ones, as discussed later on). For 'non-sorted' reference states, the reference position of a fluid parcel is then obtained as the implicit solution of the so-called level of neutral buoyancy (LNB) equation, $\rho = \rho_r(Z_r,t)$, which holds the key to the mathematical study of the reference state properties even when the reference state is not explicitly known \citep[e.g.,][]{tailleux13,saenzetal15}. 

\cite{shepherd93} showed that the local APE frameworks of \cite{andrews81} and \cite{hollidaymcintyre81} could be naturally explained in the context of Hamiltonian theory by the suitable introduction of `Casimirs', and introduced the term ``pseudo-energy'' to refer to the sum of kinetic energy plus available potential energy density, allowable in principle to account for momentum constraints as well, which was later explored by \citet{codobanshepherd03}. Shepherd's pseudo-energy was in turn connected to the concept of exergy by \cite{kucharski97} as measuring the departure from a state of mechanical equilibrium with a vertically varying temperature profile (instead of the uniform temperature $T_0$ characterizing global thermodynamic equilibrium), thus establishing the formal equivalence between the different concepts. Using this definition, \citet{kucharskithorpe00} then presented the local distributions of the zonal mean-based APE and conversion terms in a primitive-equation model. However, use of the exact local APE framework for the study of atmospheric energetics has remained limited so far.  

This paper aims to advocate the use of the local APE framework in the atmosphere as a useful tool for interpreting regional dynamics. It will 1) summarize the advantages of the local framework (Eq. \ref{EaEq}) over the Lorenz definition (Eq. \ref{Eq2}), and 2) present the first three-dimensional view of the distribution and budgets of the eddy and mean APE components in observation-based data. 
More specifically, Section 2 introduces the precise formulation of the APE, its mean and eddy components and their evolution equations. Section 3 uses ERA-Interim December-February (DJF) reanalysis data (\citealp{kallberg05}) to compare the Lorenz APE and its approximations to the exact locally derived APE when globally integrated. Section 3 also reveals the three-dimensional spatial distributions and budgets of mean and eddy local APE components which, to the authors' knowledge, has not been shown before. Section 4 summarizes and discusses the findings and their significance to the energetics community. The following analysis is limited to the dry (one-component) atmosphere, which still preserves the general features of the large scale dynamics (\citealp{pavan99}).

\section{Local APE for a Hydrostatic Dry Atmosphere}

The derivation of a local principle for the APE of a dry hydrostatic atmosphere was previously addressed by \citet{shepherd93} in the context of Hamiltonian theory. His Eq. 8.1 (using his notations) for the pseudo-energy is given by:
	\begin{equation}
	{\cal A} = \int \left \{ 1/2 g^{-1} |{\bf v}_h|^2 - \int_0^{\theta-\theta_0} c_p \left [ \Pi ({\cal P}(\theta_0 + \hat{\theta} ) - \Pi ( {\cal P}(\theta_0)) \right ] {\rm d} \hat{\theta} \right \} {\rm d}{\bf x}_h {\rm d}p ]
	\end{equation}
	where $\Pi(p)$ is the Exner function, and ${\cal P}(\theta)$ his notation for the reference pressure profile viewed as a function of potential temperature $\theta$. 

The main aim of this section is: 1) to present an alternative and arguably simpler approach that is more directly connected to the work of buoyancy forces, similar to the expressions for APE density obtained for a fully compressible non-hydrostatic fluid by \citet{andrews81} and for Boussinesq fluids by \citet{hollidaymcintyre81} and \citet{tailleux13}; and 2) to show how to obtain an exact and rigorous partition of the APE density into mean and eddy components for arbitrary Reynolds averaging operators for the study of eddy-mean flow interactions, which extends and refines previous related work by \citet{scottiwhite2014} derived in the context of the Boussinesq equations for a fluid with a linear equation of state.

\subsection{Construction and Basic Properties}
In the following, we use a pedagogical approach to construct the local APE and show its connection to the kinetic energy. To do so, we use an elementary manipulation of the horizontal momentum, hydrostatic, mass conservation and thermodynamic equations written in the following form:
\begin{equation}
\frac{D{\bf V}}{Dt} + f {\bf k} \times {\bf V} + \nabla_p 
(\Phi - \Phi_r) = {\bf F} ,
\label{hor_momentum}
\end{equation}
\begin{equation}
\frac{\partial (\Phi-\Phi_r)}{\partial p} = - [\alpha(\theta,p)-\alpha(\theta_r(p,t),p) ] ,
\label{ver_momentum}
\end{equation}
\begin{equation}
\nabla_p \cdot {\bf V} + \frac{\partial \omega}{\partial p} = 0,
\label{lcontinuity}
\end{equation}
\begin{equation}
c_p \frac{D\theta}{Dt} = \frac{\theta}{T} Q ,  
\end{equation}
where ${\bf V} = (u,v)$ is the horizontal velocity, $\omega = Dp/Dt$ is the vertical pressure velocity, $\Phi$ is the geopotential, $f$ is the Coriolis parameter, ${\bf F}$ a horizontal frictional force, 
and $\theta_r(p,t)$ is a time-dependent reference potential temperature profile whose computation is described in Appendix (\ref{App_B}). 

The equation of state for the specific volume can be written as $\alpha = RT/p = R \theta \Pi/p$, where $\Pi = (p/p_0)^{R/c_p}$ is the Exner function. For reasons that will be clarified below, it is also useful to regard the specific volume as the partial derivative of specific enthalpy $h=c_p T = c_p \Pi \theta$ at constant $\theta$, that is
$$
\alpha = \left . \frac{\partial h}{\partial p} \right |_{\theta} = c_p \theta \frac{\partial \Pi}{\partial p}.
$$

\par

An evolution equation for kinetic energy can be obtained in the usual way by multiplying the horizontal momentum Eq. \ref{hor_momentum} by ${\bf V}$, and adding it to the hydrostatic Eq. \ref{ver_momentum} multiplied by $\omega$:
\begin{equation}
\frac{D}{Dt} \frac{{\bf V}^2}{2} + \nabla_h \cdot ( \Phi' {\bf V} ) + \frac{\partial (\omega \Phi')}{\partial p} = - [\alpha(\theta,p)-\alpha (\theta_r(p,t),p) ] \frac{Dp}{Dt} + {\bf F}\cdot {\bf V}.
\label{kinetic_energy_equation}
\end{equation}
The term responsible for the conversion between kinetic energy and available potential energy is the first term on the right-hand side that is proportional to $Dp/Dt$. Here, the key is to recognize that this term can be naturally expressed in terms of the total derivative of the following quantity:
\begin{equation}
E_a(\theta,p,t) = \int_{p_r}^p \left [ \alpha(\theta,p') - \alpha(\theta_r(p',t),p') \right ] \,{\rm d}p' ,
\label{EaEq2}
\end{equation}
which we will take as our definition of local APE density, where the reference pressure $p_r=p_r(\theta,t)$ is defined to satisfy the level of neutral buoyancy (LNB) equation $\alpha(\theta,p_r) = \alpha(\theta_r(p_r,t),p_r)$, similarly to \citet{tailleux13}. It is easy to verify that the LNB equation is equivalent to the equation $\theta_r(p_r,t) = \theta$, due to the special form of the equation of state for a perfect gas. Next, the total derivative of $E_a$ can be written as:
\begin{equation}
\begin{split}
\frac{DE_a}{Dt} =  (\alpha - \alpha_r ) \frac{Dp}{Dt} + \int_{p_r}^p \frac{\partial \alpha}{\partial \theta}{\rm d}p' \frac{D\theta}{Dt} - \int_{p_r}^p \frac{\partial \alpha_r}{\partial t}\,{\rm d}p' = \delta \alpha \, \omega + \Upsilon \,Q - \chi , 
\end{split}
\label{ape_evolution}
\end{equation}
where we defined $\alpha_r = \alpha(\theta_r(p,t),t)$ for convenience. Using the fact that $\alpha = R \theta \Pi/p = c_p \theta \partial \Pi/\partial p$, it follows that we can write:
\begin{equation}
\int_{p_r}^p \frac{\partial \alpha}{\partial \theta} {\rm d}p' 
\frac{D\theta}{Dt} = c_p [\Pi(p) - \Pi(p_r)] \frac{D\theta}{Dt} = c_p \left ( \frac{T-T_r}{\theta} \right ) \frac{D\theta}{Dt} = \left ( \frac{T-T_r}{T} \right ) Q,
\end{equation}
which defines the thermal efficiency $\Upsilon$ as
\begin{equation}
\Upsilon = \frac{\Pi(p)-\Pi(p_r)}{\Pi(p)} = 1 - (p_r(\theta,t)/p)^{\kappa} = 
\frac{T-T_r}{T} , 
\end{equation}
which is the same as was previously derived by \citet{lorenz55b}, and is generally denoted by $N$ in the atmospheric APE literature.

We also defined an additional diabatic term due to temporal changes in the reference state:
\begin{equation}
\chi =  \int_{p_r}^p \frac{\partial \alpha_r}{\partial t}{\rm d}p' 
= \int_{p_r}^p \frac{R\Pi(p')}{p'} \frac{\partial \theta_r}{\partial t}(p',t){\rm d}p' .
\label{chieqn}
\end{equation}
Note that $\chi=0$ when the reference state is chosen to be independent of time. By combining Eq. \ref{ape_evolution} with Eq. \ref{kinetic_energy_equation}, the following evolution equation for the total mechanical energy (kinetic energy plus available potential energy) is obtained:
\begin{equation}
\frac{D(E_k+E_a)}{Dt} + \nabla_p \cdot ( \Phi' {\bf V} ) + \frac{\partial (\omega \Phi')}{\partial p} = {\bf F}\cdot {\bf V} +  \left ( \frac{T-T_r}{T} \right ) Q - \int_{p_r}^p \frac{\partial \alpha_r}{\partial t}{\rm d}p' . 
\end{equation}
We make the following remarks:
\begin{itemize}
	\item Our Eq. \ref{EaEq2} for the local APE density has a clear interpretation in terms of the work against buoyancy forces, similarly as in \citet{hollidaymcintyre81}, \citet{andrews81} and \citet{tailleux13}. In fact, its expression is identical to that used for estimating the Convective Available Potential Energy (CAPE) in conditionally unstable soundings \citep[e.g.,][]{emanuel94}, the only difference being the use of an arbitrary reference profile, $\alpha_r(\theta,p)$, instead of one defined by a sounding;   
	\item Eq. \ref{EaEq2} is positive definite. Its expression in the small amplitude is most conveniently expressed by regarding the reference potential temperature profile $\theta_r$ as a function of the Exner function rather than of pressure. By using the LNB equation $\theta = \theta_r(p_r,t)$, it is easy to establish that:
	$$
	E_a = \int_{p_r}^p c_p \frac{\partial \Pi}{\partial p}(p') \left [ \theta_r(p_r,t) - \theta_r(p,t) \right ]{\rm d}p' 
	$$
	\begin{equation}
	= - c_p \int_{\Pi_r}^{\Pi} \int_{\Pi_r}^{\Pi'} \frac{\partial \theta_r}{\partial \Pi}(\Pi",t) {\rm d} \Pi" {\rm d}\Pi' 
	\approx - c_p \frac{\partial \theta_r}{\partial \Pi}(\Pi_r,t) 
	\frac{(\Pi-\Pi_r)^2}{2} . 
	\end{equation}
	This small-amplitude limit for $E_a$ appears to be new as well, and is simpler than the ones obtained previously \citep[e.g.,][]{shepherd93}.
	\item An important feature of Eq. \ref{ape_evolution} is the presence of a nonlocal term proportional to $\partial \theta_r/\partial t$ that is absent from Lorenz global construction, but which can occasionally be important locally.
	\item As in \citepossess{shepherd93} expression, Eq. \ref{EaEq2} does not require the temperature reference profile to be necessarily obtained from an adiabatic re-arrangement of fluid parcels.
\end{itemize}


%

\subsubsection*{Evolution of the reference temperature profile}

As discussed above, the reference temperature profile is linked to the actual temperature through the LNB equation $\theta_r(p_r,t) = \theta$. This property can be exploited to derive an evolution equation for $\theta_r(p,t)$ in terms of the isentropic averaged diabatic heating. Indeed, the relation implies $D\theta_r(p_r,t)/Dt = D\theta/Dt = \theta Q/T$. Expanding the latter relation yields: 
$$
c_p \frac{D}{Dt} \theta_r(p_r,t) 
=    c_p \left ( \frac{\partial \theta_r}{\partial t} + \omega_r \frac{\partial \theta_r}{\partial p_r} \right ) = \frac{Q}{\Pi(p)} = \frac{\theta Q}{T} = \theta_r(p_r,t) \frac{Q}{T} 
$$
where $\omega_r = Dp_r/Dt$. It follows that by averaging on constant $p_r$ surfaces, one obtains:
$$ 
c_p \frac{\partial \theta_r}{\partial t} (p_r,t) = \overline{Q/\Pi}^{p_r} = \theta \overline{\left (\frac{Q}{T} \right )}^{p_r} ,
$$
where the overbar denotes averaging along a $p_r={\rm constant}$ surface, which at constant time coincides with an isentropic surface. This shows that the $\chi$ term in Eq. \ref{chieqn} is diabatic, and relates to the heating of the reference state.

\subsection{Separation into mean and eddy components}

The separation of energy reservoirs into mean and eddy components traditionally relies on the introduction of a Reynolds average $\overline{(.)}$ satisfying the properties for any scalar quantity $Q$, 1) $Q = \overline{Q}+ Q'$, 2) $\overline{Q'}= 0$ and 3) $\overline{\overline{Q}}=\overline{Q}$. In the context of studies of the atmospheric and oceanic energy cycles, zonal averaging has been primarily used for atmospheric studies \citep[e.g.,][]{lorenz55}, whereas temporal averaging is more characteristic of oceanic studies \citep[e.g.,][]{vonstorchetal12,zemskovaetal15}. Other important forms of averaging are ensemble average and Lanczos filtering (although the latter does not fully satisfy the classical properties of a Reynolds average). 

For a quadratic quantity such as kinetic energy, regardless the average chosen, yields a simple mean/eddy decomposition of the form $\overline{E_k}=E_k^m + E_k^e$ with $E_k^m = \overline{{\bf V}}^2/2$ and $E_k^e = \overline{{\bf V}'^2}/2$. What distinguishes APE density from kinetic energy is that it is not naturally a quadratic quantity. Thus it requires a different approach when splitting it into mean and eddy components. To that end, it is useful to introduce a non-conventional 'mean' pressure $\hat{p}_r \ne \overline{p}_r$ that differs from its Reynolds average, but one that is nevertheless unaffected by the averaging operator so that $\overline{\hat{p}_r} = \hat{p}_r$. In this study, $\hat{p_r}$ is found using $\overline{\theta}_r(\hat{p}_r,t)=\overline{\theta}$, and is a function of time and the spatial coordinates (mirroring the dimensions of $\overline{\theta}$). Note here that similar ideas enter the definition of various nonstandard 'mean' fields in the theory of the so-called thickess-weighted averaged (TWA) equations \citep[e.g.,][]{young12}. As a result, we can write
\begin{equation}
E_a = \int_{p_r}^{\hat{p}_r} \frac{R\Pi(p')}{p'}(\theta-\theta_r(p',t)) {\rm d}p' + \int_{\hat{p}_r}^p \frac{R\Pi(p')}{p'} (\theta-\theta_r(p',t)) {\rm d}p'
\end{equation}
so that taking the average enables the mean and eddy terms ($\overline{E}_a=E_a^m+E_a^e$) to be written as:
\begin{equation}
E_a^{m}=\int^p_{\hat{p}_r} \alpha(\overline{\theta},p^{\prime})-\alpha(\overline{\theta}_r(p^{\prime},t),p^{\prime}) dp^{\prime}, \label{EaEqM}
\end{equation}
\begin{equation}
E_a^{e}=\overline{\int^{\hat{p}_r}_{p_r} \alpha(\theta,p^{\prime})-\alpha(\theta_r(p^{\prime},t),p^{\prime}) dp^{\prime}}. \label{EaEqE}
\end{equation}

\subsubsection*{Evolution equations for the mean and eddy APE} 

Evolution for the mean APE obtained by taking material (Lagrangian) derivative of Eq. \ref{EaEqM}.
$$
\frac{D_ME_a^m}{Dt} = \left ( \alpha(\overline{\theta},p) - \alpha(\overline{\theta}_r(p,t),t) \right ) \overline{\omega} + \int_{\hat{p}_r}^p c_p \frac{\partial \Pi}{\partial p}(p'){\rm d}p' \frac{D_M \overline{\theta}}{Dt} - c_p \int_{\hat{p}_r}^p \frac{\partial \Pi}{\partial p}(p')\frac{\partial \overline{\theta}_r}{\partial t}\,{\rm d}p' 
$$
\begin{equation}
= \left ( \alpha(\overline{\theta},p) - \alpha(\overline{\theta}_r(p,t),t) \right ) \overline{\omega} + c_p \left [\Pi(p) - \Pi(\hat{p}_r) \right ] \frac{D_M \overline{\theta}}{Dt} - c_p \int_{\hat{p}_r}^p \frac{\partial \Pi}{\partial p}(p')\frac{\partial \overline{\theta}_r}{\partial t}\,{\rm d}p' \label{initA}
\end{equation}
where
\begin{equation}
\frac{D_M}{Dt} = \frac{\partial}{\partial t} + \overline{u}\frac{\partial}{\partial x}+ \overline{v} \frac{\partial}{\partial y} + \overline{\omega}\frac{\partial}{\partial p} 
\end{equation}
denotes advection by the mean flow. The above equations depend on the Reynolds averaged thermodynamic equation for potential temperature, which is easily shown to be:
\begin{equation}
\frac{D_M \overline{\theta}}{Dt} = - \nabla \cdot \overline{{\bf v}'\theta'} - \frac{\partial \,\overline{\omega'\theta'}}{\partial p} + \frac{\overline{\theta}}{\overline{T}} \frac{\overline{Q}}{c_p} .
\end{equation}
Note that the latter equation exploits the very special property that $\Pi(p) = \overline{T}/\overline{\theta} = T/\theta$. We can also define a mean reference temperature as $\hat{T}_r = \Pi(\hat{p}_r) \overline{\theta}$, which is not a Reynolds average, hence denoted by a hat. We can define a mean thermodynamic efficiency as the following equivalent mathematical relations:
\begin{equation}
\hat{\Upsilon} =  \frac{\overline{T}-\hat{T}_r}{\overline{T}} = 1 - \frac{\Pi(\hat{p}_r)}{\Pi(p)} = \frac{p^{\kappa} - \hat{p}_r^{\kappa}}{p^{\kappa}}
\label{ThermalEfficiency}
\end{equation}
The second term of Eq. \ref{initA} can therefore be rewritten as 
$$
c_p \left [ \Pi(p) - \Pi (\hat{p}_r ) \right ] \frac{D_M\overline{\theta}}{Dt} = \hat{\Upsilon}\, \overline{Q} - \nabla \cdot [ (\Pi(p)-\Pi(\hat{p}_r) \overline{{\bf u'}\theta'} ] + \overline{{\bf u}'\theta'}\cdot \nabla [\Pi(p)-\Pi(\hat{p}_r)]
$$
where the quantity $\Pi(p)-\Pi(\hat{p}_r) = \hat{\Upsilon} \overline{T}/\overline{\theta}$.
The evolution equation for the mean APE can therefore be written in the form:
\begin{equation}
\frac{\partial E_a^m}{\partial t} = \underbrace{-\overline{\bf u}\cdot \nabla E_a^m} _{\rm advection
} + \underbrace{\overline{\delta \alpha}\,\overline{\omega}}_{\rm C[E_k^m\rightarrow E_a^m]} - \underbrace{c_p \nabla \cdot \left ( \overline{{\bf u}'\theta'} \frac{\overline{T}\hat{\Upsilon}}{\overline{\theta}} \right ) + c_p \overline{{\bf u}'\theta'}\cdot \nabla \left ( \frac{\overline{T}\hat{\Upsilon}}{\overline{\theta}} \right )}_{\rm C[E_a^m\rightarrow E_a^e]} \underbrace{- \hat{\chi} + \hat{\Upsilon}\overline{Q}} _{\rm diabatic}
\label{Mean_APE}
\end{equation}
where
\begin{equation}
\hat{\chi} = \int_{\hat{p}_r}^p \frac{\partial \Pi}{\partial p}(p') \frac{\partial \overline{\theta}_r}{\partial t} \,{\rm d}p' .
\end{equation}
Note that again $\hat{\chi}=0$ when the reference state is chosen to be independent of time. The physical interpretation of the terms on the RHS of Eq. \ref{Mean_APE} is indicated. Namely, these terms are mean advection of the mean APE, conversion between the mean APE and mean kinetic energy $\rm (C[E_k^m\rightarrow E_a^m])$, conversion between the mean APE and eddy APE $\rm (C[E_a^m\rightarrow E_a^e])$, and a diabatic heating term. The $\rm C[E_k^m\rightarrow E_a^m]$ conversion is equivalent to that of the QG Lorenz definition. The first of the $C\rm [E_a^m\rightarrow E_a^e]$ terms vanishes under global integration. The second term of the conversion is similar to the QG Lorenz conversion, though it includes an additional component which becomes important under large static stability, as is demonstrated below.
$$
\rm{C[E_a^m\rightarrow E_a^e]_2}= c_p \overline{{\bf u'}\theta'} \cdot \nabla \left ( \frac{\overline{T}}{\overline{\theta}} \hat{\Upsilon} \right ) = c_p \overline{{\bf u'}\theta'}\cdot \nabla [ \Pi(p) - \Pi(\hat{p}_r) ]
$$
To that end, note that from the defining relation of $\hat{p}_r$, namely $\overline{\theta}_r(\hat{p}_r,t) = \overline{\theta}$, we can write:
$$
\frac{\partial \overline{\theta}_r}{\partial p} \nabla \hat{p}_r = \nabla \overline{\theta}
$$
As a result, the $\rm C[E_a^m\rightarrow E_a^e]_2$ conversion term becomes:
$$
\rm{C[E_a^m\rightarrow E_a^e]_2 }= c_p \left [ \frac{\partial \Pi}{\partial p}(p) \overline{\omega'\theta'} - \frac{\partial \Pi}{\partial p}(\hat{p}_r) \left ( \frac{\partial \overline{\theta}_r}{\partial p} \right )^{-1} \overline{{\bf u}'\theta'} \cdot \nabla \overline{\theta} \right ]
$$
$$
= \frac{R}{p_0^{\kappa}} \left [p^{\kappa-1} - \hat{p}_r^{\kappa-1} \left ( \frac{\partial \overline{\theta}_r}{\partial p} \right )^{-1} \frac{\partial \overline{\theta}}{\partial p} \right ] \overline{\omega'\theta'}
- \frac{R \hat{p}_r^{\kappa-1}}{p_0^{\kappa}} \left ( \frac{\partial \overline{\theta}_r}{\partial p} \right )^{-1} \overline{{\bf v}'\theta'}\cdot \nabla_p \overline{\theta} 
$$
$$
= \frac{R}{p_0^{\kappa}}\left [ p^{\kappa-1} - \hat{p}_r^{\kappa-1} \frac{\partial \hat{p}_r}{\partial p} \right ] \overline{\omega'\theta'} - \frac{R\hat{p}_r^{\kappa-1}}{p_0^{\kappa}} \left (\frac{\partial \overline{\theta}_r}{\partial p} \right )^{-1} \overline{{\bf v}'\theta'} \cdot \nabla_p \overline{\theta} 
$$
This term is dominated by the second term that involves the isobaric gradient of the mean temperature $\nabla_p \overline{\theta}$. The case where mean APE is converted to eddy APE corresponds to the case where $ \rm{C[E_a^m\rightarrow E_a^e]_2 }<0$. This corresponds to the case where $\overline{{\bf v}'\theta'} = - K_e \nabla_p \overline{\theta}$ is downgradient, in which case
$$
\rm{C[E_a^m\rightarrow E_a^e]_2 } \approx K_e \frac{R \hat{p}_r^{\kappa-1}}{p_0^{\kappa}} \left ( \frac{\partial \overline{\theta}_r}{\partial p} \right )^{-1} |\nabla_p \overline{\theta}|^2 < 0.
$$
The resulting expression is somewhat different from the classical Lorenz expression, in that there is now a contribution from the vertical heat flux in the expression, which is small for a stable stratification, but which can become large when static stability of the mean profile $\partial \overline{\theta}/\partial p>0$, which avoids the cancellation. This new term was previously noted by \citet{zemskovaetal15}, and is one of the novelty offered by the finite amplitude framework. 

We now turn to the derivation of an evolution equation for the eddy APE. This is obtained by subtracting the evolution equation of the mean APE density from the mean of the total APE equation:
\begin{equation}
\frac{D_M E_a^e}{Dt} = \frac{D_M \overline{E_a}}{Dt} - \frac{D_M E_a^m}{Dt}
\end{equation}
where
\begin{equation}
\frac{D_M \overline{E_a}}{Dt} = \overline{{\frac{DE_a}{Dt}}} - \overline{{\bf u}'\cdot \nabla E_a'} .
\end{equation}
Given that we can write the evolution equation for the total APE density as
\begin{equation}
\frac{DE_a}{Dt} = \delta \alpha \, \omega + \Upsilon Q - \chi ,
\end{equation}
it follows that the mean is given by
$$
\overline{\frac{DE_a}{Dt}} = \overline{\delta \alpha}\,\overline{\omega} + \overline{\delta \alpha' \,\omega'} + \overline{\Upsilon}\,\overline{Q} + \overline{\Upsilon' Q'} - \overline{\chi} 
= \frac{D_M \overline{E}_a}{Dt} + \nabla \cdot ( \overline{{\bf u}'E_a'} ) . 
$$
which implies
\begin{equation}
\frac{D_M (E_a^m + E_a^e)}{Dt} =  \overline{\delta \alpha}\,\overline{\omega} + \overline{\delta \alpha' \,\omega'} + \overline{\Upsilon}\,\overline{Q} + \overline{\Upsilon' Q'} - \overline{\chi} 
- \nabla \cdot ( \overline{{\bf u}'E_a'} ) .
\end{equation}
Note that we have 
$$
\delta \alpha =  c_p \frac{\partial \Pi}{\partial p}(p) [\theta - \theta_r(p,t) ]
$$
hence
$$
\overline{\delta \alpha} = c_p \frac{\partial \Pi}{\partial p}(p) [ \overline{\theta} - \overline{\theta}_r(p,t)]
$$
A difficulty arises with the partitioning of the thermal efficiency into mean and eddy components. Indeed, the thermal efficiency is defined by
$$
\Upsilon = \frac{\Pi(p) - \Pi(p_r)}{\Pi(p)} 
$$
However, because
$$
\overline{\Pi(p_r)} \ne \Pi(\hat{p}_r) ,
$$
subtracting the equation for the mean APE density yields
$$
\frac{\partial E_a^e}{\partial t} = \underbrace{-\overline{{\bf u}}\cdot \nabla (E_a\lna{^e})}_{\rm mean \, advection} - \underbrace{\nabla \cdot ( \overline{{\bf u}'E_a'} )}_{\rm eddy \, advection}  + \underbrace{\overline{\delta \alpha' \omega'}}_{\rm C[E_k^e\rightarrow E_a^e]} + \underbrace{c_p \nabla \cdot \left ( \overline{{\bf u}'\theta'} \frac{\overline{T}\hat{\Upsilon}}{\overline{\theta}} \right ) -c_p \overline{{\bf u}'\theta'}\cdot \nabla \left ( \frac{\overline{T}\overline{\Upsilon}}{\overline{\theta}} \right )}_{\rm C[E_a^m\rightarrow E_a^e]} 
$$
\begin{equation}
+    \underbrace{\overline{\Upsilon'Q'} + (\overline{\Upsilon}-\hat{\Upsilon}) \overline{Q} + \hat{\chi} - \overline{\chi}  }_{\rm diabatic}
\label{eddy_APE}
\end{equation}
The nature of the terms is again indicated. In particular we have the mean advection of the eddy APE, eddy advection of the total APE, conversion between the eddy APE and eddy KE ($\rm C[E_a^e\rightarrow E_k^e]$; equivalent in the Lorenz formulation), the $\rm C[E_a^m\rightarrow E_a^e]$ conversion, and diabatic terms due to the parcel heating and due to the environmental heating. Note the presence of additional small terms that arise from the difference between the non-conventional mean and standard Reynolds mean of some variables.

\section{Basic Illustrations} 
The main aim of this section is to illustrate the usefulness of the local framework. The first part focuses on the comparison between the local APE framework and the classical APE formulations proposed by \citet{lorenz55}, in order to demonstrate their equivalence and that the local APE provides more accurate estimates of the global APE than the commonly used OG Lorenz approximation. Then, we present the three-dimensional view of the local APE, its eddy and mean components as well as their budgets. This will reveal the zonally asymmetric distribution of the APE components as well as its usefulness in studying the spatio-temporal variability of the atmospheric circulation. 

\subsection{Global Values and Connection to Lorenz APE}

This section compares the globally integrated local APE density to the exact Lorenz APE (Eq. \ref{eq1}) and its QG approximations (Eq. \ref{Eq2}). Two datasets from the ERA-Interim \citep{kallberg05} archive were used, one with isobaric and one with isentropic vertical coordinates, both 6 hourly and spanning between years 1979-2016. For illustrative purposes, only data from 1 January of each year were selected, therefore selecting sets of data samples that are independent in time on daily to seasonal timescales. The reference state of the local APE was calculated by adiabatically rearranging parcels in an ascending order, i.e. by sorting all parcels based on their potential temperature using the Quicksort algorithm at each timestep. This makes the reference state equivalent to the reference state of Lorenz APE, so theoretically the exact Lorenz APE should equal to the globally integrated local APE \citep{andrews81}. For comparison, local APE calculated using a reference state that is an average potential temperature on isobaric surfaces is also displayed (see Appendix \ref{App_B} for more detail). Computation the Lorenz APE is somewhat more efficient. On a standard personal computer 20 timesteps took 6 seconds or less for the Lorenz APE diagnostics, 25 seconds for the local APE using the isobaric $\theta_r$ and 2min for the local APE using the Quicksort $\theta_r$.

\begin{figure}[t]
	\begin{center}
		\includegraphics[trim= 0mm 0mm 0mm 0mm,clip,width=0.35\textwidth]{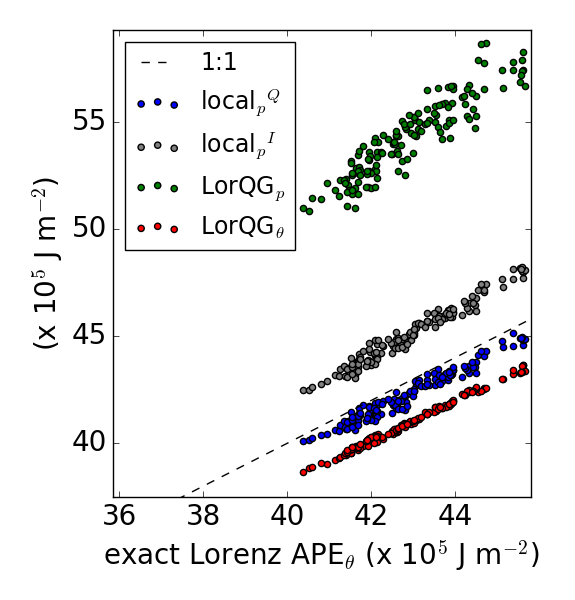} 
		\caption{Globally integrated local APE (in blue calculated using the Quicksort $\theta_r$, and in gray using the isobarically averaged $\theta_r$) compared to the exact Lorenz APE on isentropic surfaces (x-axis), and the QG Lorenz approximations on isentropic (red) and isobaric (green) surfaces. The 1:1 line is included (dashed). Era-Interim data of 1 January (4 time steps 6 h apart) of years 1979-2016. }\label{GlobIntegrated}
	\end{center}
\end{figure}

Fig. \ref{GlobIntegrated} shows that the local APE on isobaric surfaces is slightly lower than the exact Lorenz APE evaluated on isentropic surfaces. Because the minimum and maximum values of the isobaric and isentropic surfaces do not exactly match the maximum and minimum values of the variable pressure and potential temperature in the respective reanalysis datasets, an exact match between the isobarically-based local APE and isentropically-based exact Lorenz APE is not necessarily expected. Nevertheless, the local APE is the closest match to the exact Lorenz APE, better than the QG Lorenz approximation on isentropic surfaces and substantially better than the QG Lorenz approximation on isobaric surfaces (the latter being the most commonly used diagnostic for the APE). 

The QG approximation of Lorenz APE is often studied with respect to the Lorenz cycle, where both kinetic energy and QG APE on pressure surfaces are split into their mean and eddy components. The four resulting evolution equations (one for each $E_a^m$,$E_k^m$,$E_a^e$, and $E_k^e$) form a closed system in the absence of diabatic and frictional processes, which makes the system (referred to as the Lorenz cycle) an attractive theory for studying energy exchanges. 

\begin{figure}[t]
	\begin{center}
		\includegraphics[trim= 0mm 0mm 0mm 0mm,clip,width=0.80\textwidth]{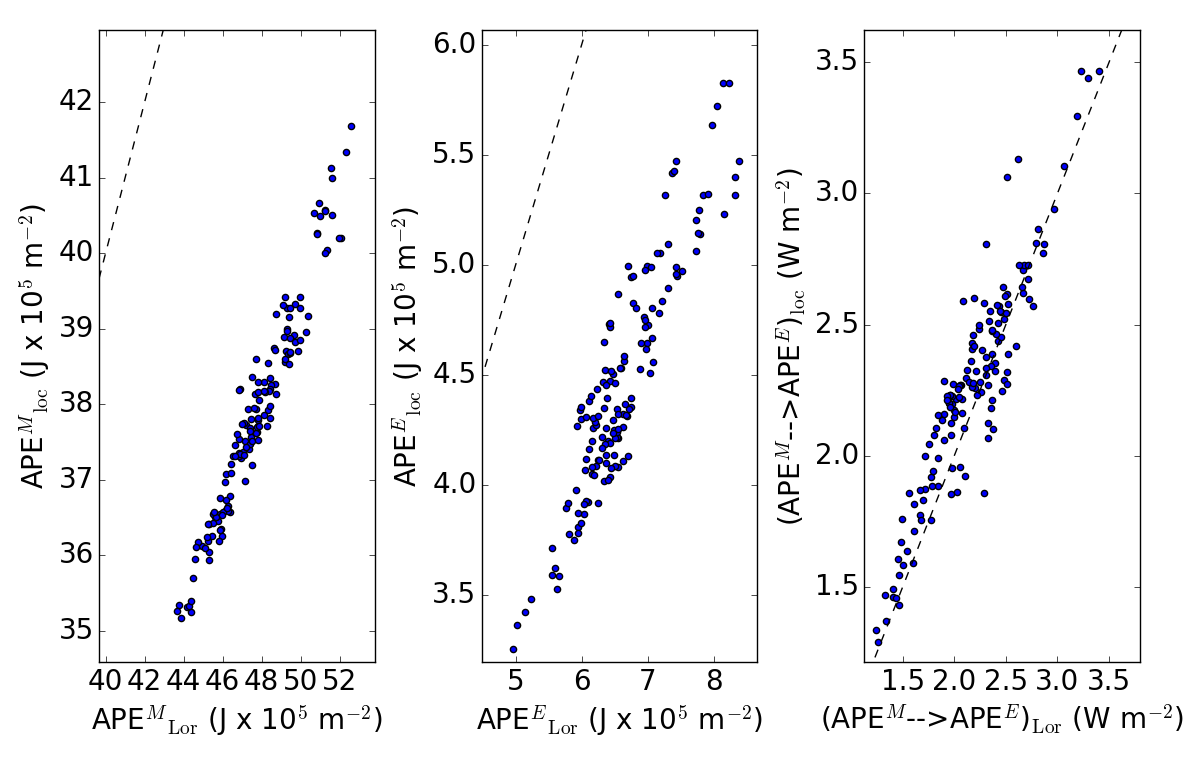} 
		\caption{Lorenz QG APE comared to the globally integrated local APE, with both being evaluated on isentropic surfaces and split into their mean (left) and eddy (middle) components. The conversion between the mean and eddy APE for the two frameworks is also shown (right). The 1:1 line is included (dashed). The data are from 1 January (4 time steps 6 h apart) of years 1979-2016 in the ERA-Interim reanalysis. }\label{GlobIntegratedCA}
	\end{center} 
\end{figure}

It is apparent from Eq. \ref{Mean_APE} and \ref{eddy_APE} that the local eddy and mean APE density equations can be used with the mean and eddy kinetic energy equations (both of which are already of a local nature), in order to obtain a local and exact version of the Lorenz cycle (i.e. a system of the four evolution equations that is closed under adiabatic conditions). When globally integrated the two energy cycles are equivalent, apart from small differences in formulation of three terms: mean APE, eddy APE and the $\rm C[E_a^m\rightarrow E_a^e]$ conversion. These three terms are compared for the QG Lorenz and globally integrated local frameworks in Fig. \ref{GlobIntegratedCA}. 

Both eddy and mean APE components are overestimated by the QG approximation, corresponding to the total APE being larger. The conversion term is of a similar magnitude with some spread, resulting from the QG approximation being less accurate under large static stability. However, we have found that there is no simple linear relationship between static stability and the difference between the two conversions. 

\subsection{Spatial Distribution and Variance of APE Components}

This section focuses on the three-dimensional structure of the eddy and mean APE components and their inter-annual variability. We focus on the winters (December-February) of years 1979-2016 in daily-averaged ERA-Interim data. The winter season was selected because during it the mid-latitudes are dominated by strong eddies which are particularly interactive with the mean large-scale circulation, and this interaction will be the focus of a forthcoming paper. 

For the same reason, we separate the APE into eddy and mean components using the 10-day Lanczos filter \citep{duchon89a}. This way the eddy component mainly corresponds to high-frequency baroclinic transients that are associated with synoptic storms, and the mean component corresponds to more slowly varying circulations, such as the mid-latitude jet \citep{hoskinsetal83,novaketal15}. Another (more technical) advantage of this separation is that it also allows investigation of APE in all three spatial dimensions (rather than only two when the zonal mean is used for the separation), as well as in time (which would not be possible had the time mean been used). Comparison between the two separation methods (i.e. using the zonal mean vs. the Lanczos filter) is shown in Appendix \ref{App_C}.

We also use the isobaric average (instead of the Quicksort method) to compute the reference potential temperature profile, because it avoids extremely high APE values at the surface due to extremely high (and potentially badly constrained) potential temperature values from the top of the atmosphere (see Appendix \ref{App_B}).

\begin{figure}[t!]
	\begin{center}
			\includegraphics[trim= 20mm 155mm 20mm 20mm,clip,width=1.\textwidth]{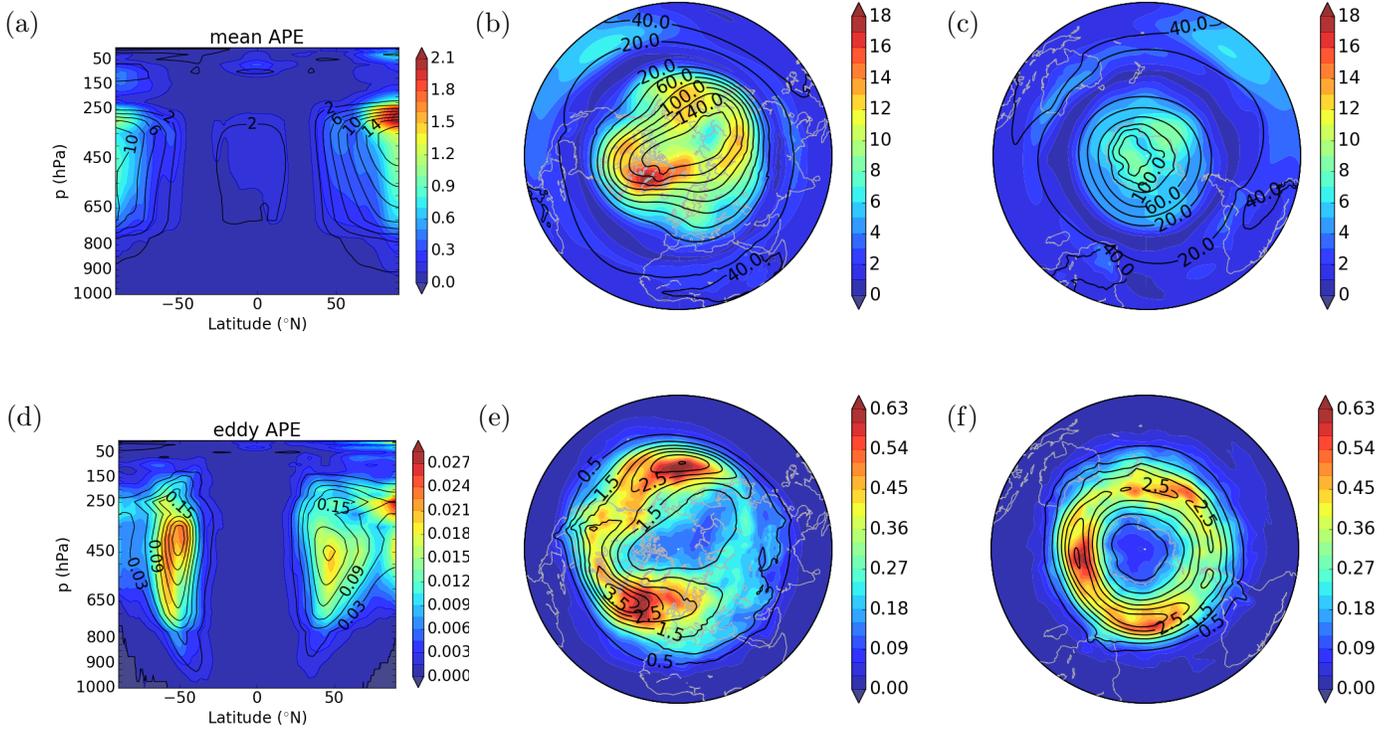}   
		\caption{Mean (top row) and eddy (bottom row) APE distributions (black contours), zonally averaged (a,d), and vertically integrated (using mass weighting) of the Northern Hemisphere (b,e) and the Southern Hemisphere (c,f). The colors refer to the annual standard deviation. The split into mean and eddy is based on the Lanczos filter, and the units are scaled to be 10$^5$ J m$^{-2}$.}\label{3dAPELanc}
	\end{center}
\end{figure}

The three-dimensional spatial distribution of the eddy and mean APE is displayed in Fig. \ref{3dAPELanc}, along with their interannual standard deviations (i.e. based on the departures of annual values from the long-term mean of all winters). In the units of J m$^{-2}$, the mean APE (top row) is most concentrated in the upper levels of high latitudes, and exhibits a minimum in the mid-latitudes with a secondary maximum in the tropics. This zonally-averaged profile is expected because a) it has been shown before \citep{kucharski97,kucharskithorpe00}, and b) by definition (since $\theta$ on average decreases continuously with latitude) the  high and low latitudes are characterized by the largest departures from the globally horizontally constant reference state (and the high latitudes are more extreme because they cover a smaller surface area). Similarly, eddy APE distribution is as expected, peaking near the upper levels of the mid-latitudes and mirroring the eddy KE \citep[e.g.,][]{kucharskithorpe00}.

The zonal asymmetries of the APE energy components are such that the mean APE follows the general structure of the mean temperature and PV fields (not shown) with maxima extending more equatorward over continents.This is especially apparent in the Northern Hemisphere. Downstream of these regions of enhanced mean APE are maxima in eddy APE, which peak over the main storm track regions over the North Atlantic, the North Pacific, and the Southern Ocean \citep[e.g.,][]{kaspischneider13}. In a thought experiment where the atmosphere could be brought to a state of zero baroclinicity (i.e. no meridional temperature gradients), the mean APE maxima can be seen as energy reservoirs that fuel the mid-latitude storm tracks between them (as is shown more explicitly in the next section). 

The standard deviations of the mean and eddy APE components are shown in colors in Figure \ref{3dAPELanc}. The highest interannual variability in the mean APE of the Northern Hemisphere is above Greenland, with secondary maxima in the North Pacific and over central northern Siberia. The Southern Hemisphere in DJF exhibits a dipole centered over the South Pole, with the stronger maximum being above the Ross Sea. Some enhanced variability is also apparent in the tropical central Pacific, where ENSO operates. The variability of the eddy APE is generally most pronounced near the central and end parts of the storm tracks.

\subsection{Thermal Efficiency}
\begin{figure}[h]
	\begin{center}
			\includegraphics[trim= 20mm 170mm 18mm 60mm,clip,width=1.\textwidth]{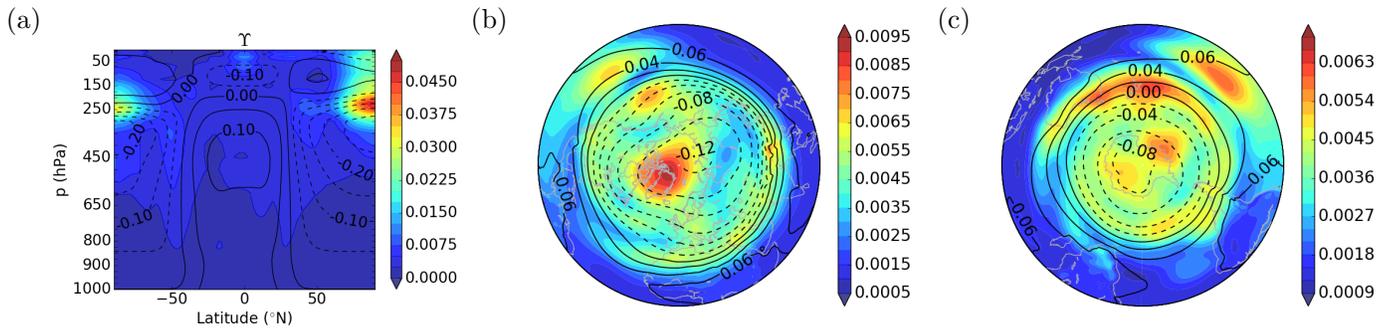}  
		\caption{Thermal efficiency, $\Upsilon$ (black contours), zonally averaged (a), and horizontally averaged (using mass weighting) of the Northern Hemisphere (b) and the Southern Hemisphere (c). The colors refer to the annual standard deviation. The efficiency is dimensionless. }\label{Efficiency}
		
	\end{center}
\end{figure}

It is of interest to investigate the thermal efficiency defined in Eq. \ref{ThermalEfficiency}, beause it is the factor that determines the sign and magnitude of the effect of a) diabatic heating on the APE generation/dissipation, and b) the APE conversion into eddy energy. This efficiency is identical to that discussed by \citep{lorenz55b} and several other authors (including \textcolor{red}{Sigmund}), and its magnitude and distribution, as shown in Fig. \ref{Efficiency}, is comparable to the previously published estimates. However, here we additionally show the full horizontal structure, as well as the interannual variability. 

Since the thermal efficiency is defined as the departure of the actual thermal state from a reference state, it is apparent that the QG Lorenz assumption of this being of a small amplitude is a poor one. The QG Lorenz APE and $\rm C[E_a^m\rightarrow E_a^e]$ terms are defined using the thermal efficiency. It is therefore unsurprising that these terms are of a somewhat different magnitude, as shown in the previous sections.

The thermal efficiency also displays high annual variability, as shown by the standard deviation in colors. The most variable regions are in the north-western Pacific, above Greenland and near the coast of West Antarctica. A cross-hemispheric wavetrain-like pattern emerges in the Central Pacific. Some of these features mimic those of the mean APE variability, and are relevant for climate sensitivity studies.

\subsection{Mean and Eddy Local APE Budgets}

\begin{figure}[t]
	\begin{center}
		\includegraphics[trim= 0mm 0mm 0mm 0mm,clip,width=1.\textwidth]{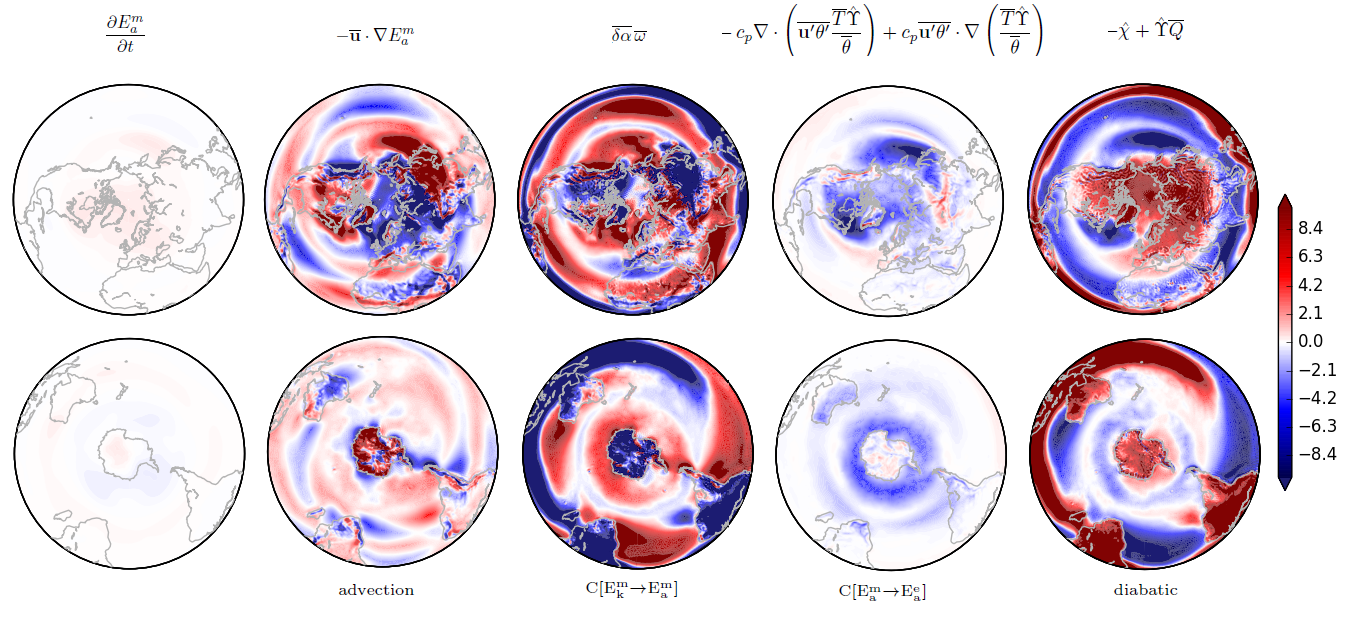} 
		\caption{Local mean APE budget (Eq. \ref{Mean_APE}) for the northern (top panel) and southern (bottom panel) Hemispheres. Units are scaled to be W m$^{-2}$. All terms are vertically integrated.}\label{budgetAM}
	\end{center}
\end{figure}

\begin{figure}[t]
	\begin{center}
		\includegraphics[trim= 0mm 0mm 0mm 0mm,clip,width=1.\textwidth]{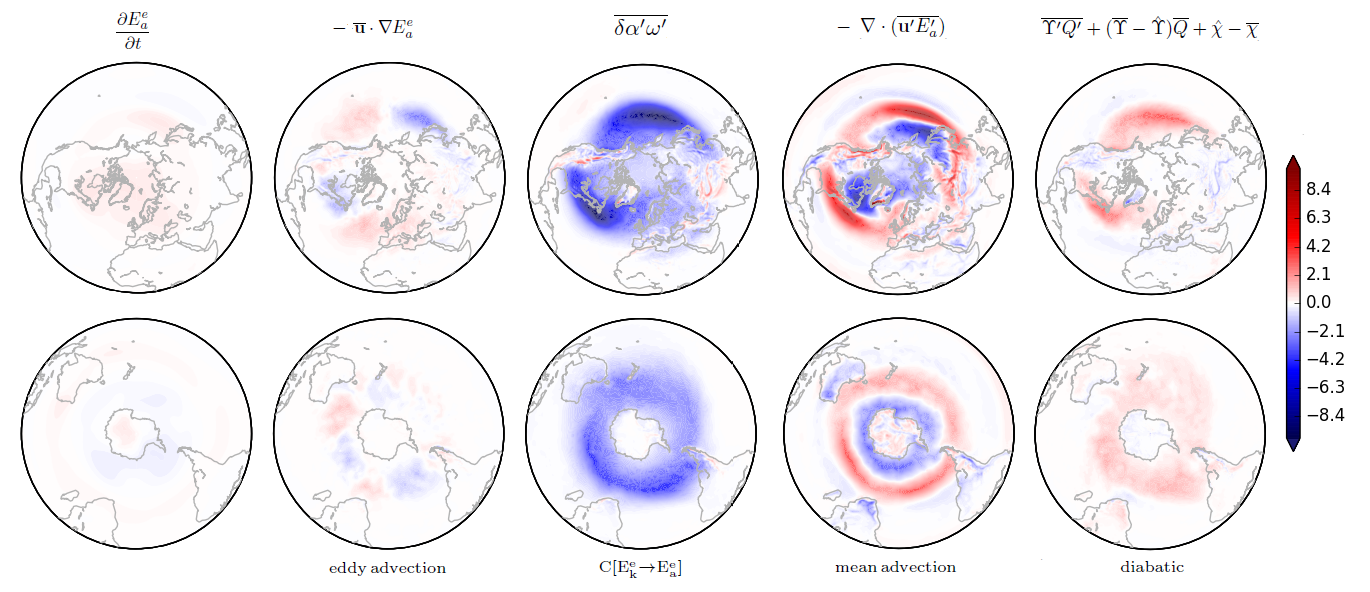} 
		\caption{Local eddy APE budget (Eq. \ref{eddy_APE}) for the northern (top panel) and southern (bottom panel) hemispheres. Units are scaled to be W m$^{-2}$. All terms are vertically integrated.}\label{budgetAE}
	\end{center}
\end{figure}

The mean and eddy local APE budgets (i.e. terms in Eq. \ref{Mean_APE} and \ref{eddy_APE}) are plotted for both hemispheres in Fig. \ref{budgetAM} and \ref{budgetAE}, respectively. The sum of the diabatic terms is calculated as a residual of the non-diabatic terms in the two evolution equations. Again, we use the 10-day Lanczos filter to separate into mean and eddy terms, but this separation produces a small leakage due to the non-commutability of the mean (i.e. $\overline{X}\neq\overline{\overline{X}}$ if the bar represents a mean derived from the Lanczos filter). This leakage is included in the residual diabatic terms. However, its magnitude was found to be small and the associated eddy and mean heating rates are comparable to those derived in previous works using different methods \citep[e.g.,][]{kallberg05}. Note that the conversion term ($\rm C[E_a^m\rightarrow E_a^e]$) is not displayed in Fig. \ref{budgetAE} of the eddy APE budget, because it is already shown in Fig. \ref{budgetAM} (only the sign would change in the eddy APE equation). The conversion and heating terms were checked against existing estimates \citep{oort64,kallberg05,AAlietal07} to ensure that the obtained values are plausible.

Turning to the first terms in both budgets, the respective tendencies of the mean and eddy APE are non-zero, even though they are averaged over time. The reader is reminded that these averages are limited to the winter season, so the non-zero values represent changes throughout that season. Though small, these changes are such that both APE components increase in the Northern Hemisphere, and decrease in the Southern Hemisphere throughout the season.  

The mean APE budget is dominated by the $\rm C[E_k^m\rightarrow E_a^m]$ conversion and the diabatic generation of the mean APE. The diabatic term was found to be almost entirely dominated by the $\overline{\Upsilon}\,\overline{Q}$ component (not shown), which itself follows observed diabatic heating and cooling rates \citep[e.g.,][]{kallberg05}. More specifically, the tropics are the main diabatic generation regions of mean APE, though there is a secondary maximum over the poles, corresponding to the large observed radiative heating an cooling rates in those regions. Both tropical heating and polar cooling increase the large-scale meridional temperature gradients, and thus the local departures from the global reference state, i.e. the mean APE. On the other hand, diabatic mean APE dissipation corresponds to these large-scale temperature gradients being reduced. This happens in the subtropics where cooling occurs over the return flow of the oceanic subtropical gyres, and within the mid-latitude storm tracks where (mostly) latent heating reduces the large-scale temperature gradients \citep{hoskinsandvaldes90,kallberg05}.

A large part of the diabatic contributions is compensated for by the $\rm C[E_k^m\rightarrow E_a^m]$ conversion, especially in the tropics. This term is relatively weak in the mid-latitudes because the circulation there is dominated by eddy motions, and the dipole in the Central Pacific mirrors the change in sign of the vertical motion of the Walker circulation. 

This conversion is often interpreted as representing the mean overturning circulation (\citealp{james94}). However, one needs to consider that this term is defined with relation to the global reference state in this study (as it is in the Lorenz framework). For example, at the poleward edge of the Hadley cell the buoyancy difference (equivalent to $\overline{\delta\alpha}$ in Equation \ref{Mean_APE}) with respect to the global reference state is positive, but the difference with respect to the immediate surroundings is negative (as is the case for parcels in thermally direct circulations, such as the Hadley Cell).  Since this region is characterized by descending motion (positive $\omega$), the $\rm C[E_k^m\rightarrow E_a^m]$ conversion in this region is positive, whereas it would be negative for a more local reference state. This demonstrates the importance of choosing the correct reference state for the study of interest. In this and Lorenz's case, the sign of the $\rm C[E_k^m\rightarrow E_a^m]$ conversion does not reflect the sign of the overturning circulation. Rather, it indicates how the local vertical motions contribute to the large-scale baroclinicity. The freedom to choose a reference state that is appropriate for the study of interest is only possible with the local framework, but not the Lorenz framework. 

The mean APE is converted into eddies ($\rm C[E_a^m\rightarrow E_a^e]$ conversion) predominantly poleward of all storm tracks, with some weak conversions on their equatorward side. This is despite the predominant diabatic generation of APE being in the tropics, suggesting that eddies preferably tap into the APE reservoir poleward of the storm track. The mean advection of the mean APE is the only term that is positive at the beginning of storm tracks, indicating that advection is crucial for supplying the mean APE to fuel storm tracks. 

Moving on to the eddy APE budget, it is apparent that conversions from mean APE ($\rm C[E_a^m\rightarrow E_a^e]$) and into eddy kinetic energy ($\rm C[E_a^e\rightarrow E_k^e]$) are the dominant terms, in agreement with observations of preferred energy flows of global energy \citep[e.g., analysis of the Lorenz cycle in][]{oort64}: $E_a^m \rightarrow E_a^e \rightarrow E_k^e$ ($\rightarrow$ friction). The  $\rm C[E_a^m\rightarrow E_a^e]$ and $\rm C[E_a^e\rightarrow E_k^e]$ conversions are located poleward of and at the location of the storm tracks, respectively.

As for the remaining terms, eddies advect the total APE to the equatorward flank of the storm tracks. A small amount of eddy APE is advected by the mean flow further downstream of the storm tracks, and a small amount is generated by sensible and latent heating from pre-existing eddies within the storm tracks (since the largest contributor is the $\overline{\Upsilon^{\prime}Q^{\prime}}$ component).  

In summary, it is evident that the classical Lorenz cycle of global energy flows is more complicated regionally. In particular, the conversion between the mean energies ($\rm C[E_k^m\rightarrow E_a^m]$) is the dominant term regionally, though it is near zero if integrated globally. It is evident that energy advection into the mid-latitudes is essential for fueling the storm tracks and that this energy is mainly supplied from high latitudes, perhaps because that is where the mean APE has to be more concentrated than in the tropics due to the spherical geometry.  

\section{Discussion and Conclusions}

In this paper, we have developed and extended \cite{hollidaymcintyre81,andrews81} local APE theory to the case of a dry hydrostatic atmosphere, and illustrated its usefulness using ERA interim data. The main new advances are: 1) a simpler mathematical expression for the APE density that is physically more revealing than that previously derived by \cite{shepherd93}; 2) accounting for diabatic effects; 3) an exact separation between mean and eddy components valid for any form of Reynolds averaging; and 4) a demonstration of the feasibility of defining reference position for fluid parcels even for non-sorted reference states. Because this formulation has seldom been used on diagnostic studies, we advocate its use by presenting its new form on isobaric coordinates, by comparing it it to the classical global definition suggested by \citet{lorenz55}, and by presenting an illustration of its usefulness for understanding the spatio-temporal variability of the large-scale circulation. 

Although the Lorenz APE definition is by far the most commonly used measure of the observed APE, we have found the following advantages if the local APE is used instead:
\begin{itemize}
	
	\item \textit{Computational feasibility versus accuracy}: Lorenz exact APE definition is based on averaging on isentropic surfaces, but datasets are rarely available in isentropic coordinates. Additionally, Lorenz APE is the difference between two large terms (the potential energy of an actual state and that of a reference state), which makes the calculations highly sensitive to small numerical errors. The Lorenz APE is therefore most commonly diagnosed as its QG approximation on isobaric surfaces, but this easily computable approximation comes at the cost of accuracy. On the other hand, one does not have to compromise with the local APE density, which is both exact for finite amplitude departures from Lorenz reference state, and easily computable from isobarically-based data. Additionally, the globally integrated local APE density is a sum of small positive definite values (instead of a difference between two large terms), which is always preferable from the computational viewpoint. Although the computation time of the local APE is four times as large as that for the Lorenz diagnostics (though the exact computation time depends on the method to calculate $\theta_r$), this remains manageable and seems a small price to pay in view of the considerably greater accuracy achieved.
	
	\item \textit{Energy conservation}: It seems useful to point out that the relevant form of APE density depends on the particular approximate set of equations used. Although Lorenz defined APE as the difference in potential energy between the actual state and that of a reference state obtained by an adiabatic re-arrangement of mass, it is not normally realized that this makes sense only for the full Navier-Stokes equations that describe a non-hydrostatic compressible atmosphere. Indeed, this is because for a hydrostatic atmosphere (as is generally considered in practice), the relevant energy quantity (which defines the Hamiltonian) is ${\bf V}^2/2+h$, where ${\bf V}$ is the horizontal velocity and $h$ the specific enthalpy \citep[e.g.,][]{shepherd93}. In other words, the standard form of energy ${\bf U}^2 + g z + e$ is not a conservative quantity for the hydrostatic primitive equations. This being said, the volume integral of the APE density for a hydrostatic atmosphere can be written as
		\begin{equation}
		\int_{V} E_a {\rm d}m = \int_{V} (h-h_r)\,{\rm d}m + \int_{V} (\Phi_r(p) - \Phi_r(p_r))\,{\rm d}m ,
		\label{integral_APE}
		\end{equation}
		and is not necessarily equal to the enthalpy difference between the actual and reference states in presence of orography, as the latter generally causes the last term on the right-hand side of Eq. \ref{integral_APE} to be non zero in general.

	\item \textit{Local diagnosis}: Some studies use the Lorenz QG approximation locally, which gives physically plausible results \citep[e.g.,][]{AAlietal07}. However, the Lorenz APE definition formally relies on the APE being defined as a global integral. This is not the case for the local APE, which can be formally and exactly defined locally. In combination with the kinetic eddy and mean energies (which are already of a local nature), the local APE can also be used to derive a local version of the Lorenz energy cycle.
	
	\item \textit{Local reference state}: As opposed to the Lorenz definition which requires the reference state to be an adiabatic re-arrangement of the actual global state, the local APE framework accommodates other reference states that can be locally defined (much like buoyancy).
	
\end{itemize}

As predicted theoretically, using reanalysis data we confirmed that the globally integrated local APE is more comparable to the exact Lorenz definition than the QG Lorenz approximation. A rather surprising discrepancy was found between the QG APE computed on insentropic surfaces and the QG APE computed on isobaric surfaces, the two of which are often assumed to be closely related \citep{lorenz55}.

We also demonstrated how the mean and eddy components of APE vary both in space and in time in 35 DJF seasons in both hemispheres, so both the winter and summer seasons were studied. We used the 10-day Lanczos filter to separate the APE into its eddy and mean components, so that eddy quantities had the characteristics of synoptic-scale eddies. Although we were not restricted to using the global state to define the local APE, we chose to do so, in order to stay in the context of the existing literature. We defined the global reference state as the isobaric average of potential temperature. The disadvantage of using a global reference state is that (as in the Lorenz definition) the atmosphere is assumed to be capable of equilibrating itself to a state of zero baroclinicity, which is clearly not something that is observed. Nevertheless, insightful results can be obtained, as long as the dependency of the APE on its reference state is considered with care. 

The local APE represents the largest temperature deviations from the global insentropic average, so the APE is especially abundant over the poles and the tropics, as shown before by \cite{kucharski97,kucharskithorpe00}. As far as we are aware, the zonally asymmetric APE distribution is shown here for the first time, and it is particularly zonally asymmetric in the Northern Hemisphere, following a similar structure to the large scale mean temperature or potential vorticity. The local APE should not be seen as the growth rate for baroclinic eddies, which depends on the meridional temperature gradients \citep{eady49,pedlosky92} rather than the departures from the isobaric mean. The eddy growth rate is maximum at the latitude of storm tracks \citep[e.g.,][]{hoskinsandvaldes90}, whereas APE is maximum in the polar and tropical regions.

Nevertheless, the budgets and the interannual variabilities of the mean and eddy APE components provide useful insights on baroclinic eddy growth and other aspects of the large-scale dynamics. For example, the illustrations within this study show the following. 
\begin{itemize}
	\item The classical studies of baroclinic eddy lifecycles (e.g. \citealp{simmonshoskins78}) have shown that the primary energy exchanges, as diagnosed by the Lorenz framework, are:
	$$
	\rm{diabatic \,\, processes} \rightarrow E_a^m \rightarrow E_a^e \rightarrow E_k^e (\rightarrow friction)
	$$
	This energy pathway is also observed for global time mean observations of the atmosphere \citep[e.g.,][]{oort64}, indicating that the global energetics are primarily governed by baroclinic instability. While this pathway also seems to exist locally within the storm tracks, it is apparent that the conversion between the mean energies often dominates despite its global average being near zero. The primary role of this conversion, which reflects ageostrophic circulation, is to compensate for a large part of the mean diabatic heating. This is not obvious from Lorenz's formulation.
	
	\item Advection terms of the mean and eddy APE are obscured in the globally integrated framework. Nevertheless, it is shown here that advection of the mean APE is essential for providing APE into (and increasing baroclinicity within) the storm tracks, rather than APE being generated diabatically in situ by processes, such as SST heating.
	
	\item The mean APE advection is primarily from latitudes poleward of the storm tracks, which may have implications on the latitudinal extent of storm tracks. It was shown in \cite{novaketal15} that the equatorward part of the North Atlantic storm track is anchored near the latitude of the subtropical jet. This can be explained by the Hadley Cell edge being anchored by a tropical energy balance \citep{mbengueschneider17b}. However, \cite{novaketal15} also find that the poleward edge of storm tracks is much more transient, which may be because the advection of cold temperatures determines the extent to which eddies can grow and propagate poleward.
	
	\item The analysis above revealed an interesting variability of APE in the upper level troposphere over Greenland. The variabilities of mean APE and thermal efficiency exhibit an inter-hemispheric wave train that emanates from the ENSO region and propagates into the higher latitudes. The wave train in the southern hemisphere reaches the Ross Sea, and the variability may be even more prominent in the winter. With Greenland and Ross Sea being predicted to experience large changes in their land and sea ice coverage \citep{,jacobsetal02,shepherdwingham07,jacobsetal11}, it is possible that these regions of main supply of mean APE into storm tracks will play an important role in mid-latitude dynamics in the future climate. 
\end{itemize}
We are keen to emphasize that these are merely a few illustrations of the usefulness of the local APE as an atmospheric diagnostic. For a thorough specific analysis of storm tracks, one may wish to optimize the choices of the reference state and separation methods into eddy and mean quantities. The choices made here are primarily to facilitate comparison with existing studies. The APE density framework can be further extended to a multicomponent fluid (\citealp{bannon05}; \citealp{tailleux13}; \citealp{pengetal15}). However, the addition of moisture would introduce the possibility of parcels possessing multiple reference states. This would impact on the magnitude of APE, most likely increase it (\citealp{lorenz79}; \citealp{pauluisheld02}; \citealp{bannon05}). \citet{pengetal15} presented an application of a positive definite definition of the moist local APE, based on using the virtual temperature in an idealized atmosphere, showing the marked difference between using the classical exergy and their APE density. However, such considerations are beyond the scope of this paper, which demonstrates the usefulness of this local framework in analyzing large-scale dynamics, and provides interesting directions for further focused research.


\begin{acknowledgment} 
This work is supported by the U.K. Natural Environment Research Council [grant number NE/M014932/1]. We would like to thank Prof. Tapio Schneider for a useful discussion on APE.
\end{acknowledgment}


 \appendix[ ] 
\subsection{Two Methods for Constructing the Reference Potential Temperature} \label{App_B}

The first method uses parcel sorting. For each time the global potential temperature is divided into parcels. For example, the ERA-Interim dataset has a resolution of 512 longitude values, 256 latitude values and 37 pressure levels, giving 4,849,664 parcels. These parcels were then sorted using the Quicksort sorting algorithm for higher numerical efficiency. In this sorted order each parcel mass was then draped across the Earth's surface yielding the height of each parcel. The result is $\theta_r$ as a function of the cumulative parcel mass (which can be readily converted to pressure). The $\theta_r$ profile for the required pressure levels of the dataset was then obtained using linear interpolation.   

The second method is using the latitudinally weighted isobaric averaging as was used by \citet{lorenz55} and others. This method is faster, as discussed in the text.

\begin{figure}[h]
	\begin{center}
			\includegraphics[trim= 20mm 170mm 20mm 60mm,clip,width=1.\textwidth]{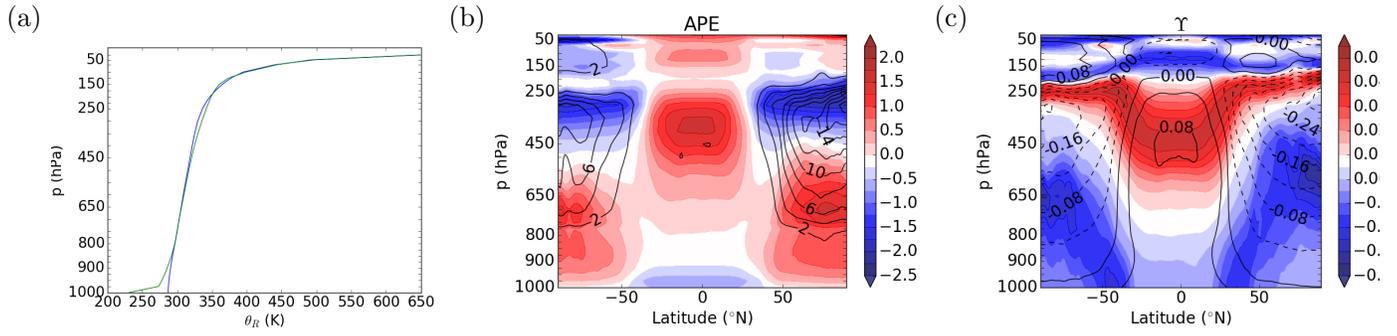}        
		\caption{Comparison of the Quicksort (black contours) isobaric averaging (shading, displayed as difference from when using the Quicksort method) methods to define the reference potential temperature profiles (a).  The resultant effect on the local APE (b, units: $10^5$ J m$^{-2}$) and thermal efficiency (c). Data used are 1 January 2000.}\label{onetimeQvIcomp}		
	\end{center}
\end{figure}

The Quicksort and isobaric averaging methods are compared here for 1 January 2000 in ERA-In. Their $\theta_r$ profiles, and zonally averaged APE and thermal efficiencies are shown in Fig. \ref{onetimeQvIcomp}. In the zonal-mean plots, the quantities derived using the isobaric averaging are shown as anomalies from those derived using the Quicksort method.

The $\theta_r$ two profiles are almost equivalent apart from the very low levels, and near the tropopause. This makes a small difference in the APE, with the Quicksort yielding a lower APE in the tropics and low-level polar regions, and higher APE in the upper level polar regions. The Quicksort method also produces a lower thermal efficiency near the tropopause and higher elsewhere. 

One could argue that the Quicksort method is more accurate. However, given the data resolution, we are only interested in the larger scale patterns in the energy and conversion terms. The smoother reference potential temperature profile is therefore still adequate and it allows a more direct comparison with the existing literature. In addition, the Quicksort method relies on all potential temperature values, including those in the highest levels, which are often not well constrained.

\subsection{Eddy and mean local APE using zonal mean and Lanczos filer}  \label{App_C}
\begin{figure}[t!]
	\begin{center}
		\includegraphics[trim= 0mm 65mm 0mm 0mm,clip,width=0.85\textwidth]{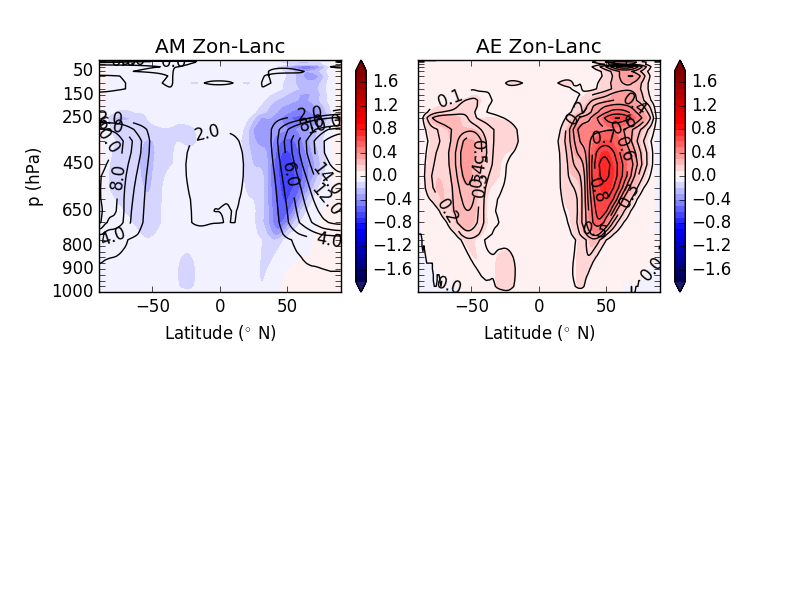} 
		\caption{Differences between the mean and eddy APE fields of the zonal mean-based and Lanczos-based  frameworks (shading). The black contours show the zonal mean-based APE components for reference. The units are scaled to be 10$^5$ J m$^{-2}$. }\label{ZonalMeansZonvLanc}
	\end{center}
\end{figure}

It is noted that the zonal mean-based framework is qualitatively similar to the Lanczos-based framework, as shown in Figure \ref{ZonalMeansZonvLanc}. As expected, a part of the mid-latitude eddies (low frequency and stationary eddies) is transferred from the eddy component to the mean component in the Lanczos-based framework. The mean APE mirrors closely the distribution of the total APE, so only the former is presented here.

\ifthenelse{\boolean{dc}}
{}
{\clearpage}


\ifthenelse{\boolean{dc}}
{}
{\clearpage}
\bibliographystyle{ametsoc}
\bibliography{references3}

%
\end{document}